\documentclass[aip,pop,reprint]{revtex4-2}
\usepackage{graphicx}
\usepackage{amsmath,amssymb,bm}
\usepackage{xcolor}
\usepackage{url}
\usepackage{hyperref}
\graphicspath{{./}{./figures/}{./Figures/}}

\begin{document}

\title{Linear-wave bound on electromagnetic energy equipartition at sub-electron scales in non-relativistic plasmas}

\author{Vivek Shrivastav}
\email{vivekshrivastav1998@gmail.com}
\affiliation{Department of Physics, Sikkim University, Gangtok-737102, India}

\author{Mani K Chettri}
\email{mkchettri8@gmail.com}
\affiliation{Department of Physics, Sikkim University, Gangtok-737102, India}

\author{Britan Singh}
\affiliation{Department of Physics, Sikkim University, Gangtok-737102, India}

\author{Hemam D. Singh}
\affiliation{Department of Physics, Netaji Subhas University of Technology, New Delhi-110078, India}

\author{Rupak Mukherjee}
%\email{rmukherjee@cus.ac.in}
\affiliation{Department of Physics, Sikkim University, Gangtok-737102, India}

\begin{abstract}
Recent Magnetospheric Multiscale (MMS) observations report approximate equality between electric and magnetic field energy spectral densities, $\varepsilon_{0} P[\delta E]/2 \approx P[\delta B]/(2\mu_{0})$, at sub-electron scales in reconnection-driven magnetotail turbulence, interpreted as relaxation toward thermodynamic equilibrium. We derive the electric-to-magnetic energy ratio from the linear polarization of kinetic Alfv\'en waves and whistler-mode waves in the two-fluid framework and show that it saturates at $\mathcal{R}_{\infty}=(V_{A}/c)^{2}(m_{i}/m_{e})(\beta_{e}/2)$ deep in the sub-electron regime. Setting $\mathcal{R}_{\infty}=1$ yields the universal threshold $V_{A}/c \gtrsim \sqrt{2/[(m_{i}/m_{e})\beta_{e}]}$, which no non-relativistic space plasma satisfies. For typical magnetotail parameters, $\mathcal{R}_{\infty}\approx 2\times 10^{-3}$, approximately 500 times below the observed value, a discrepancy rooted in the non-relativistic ordering $(V_{A}/c)^{2}\ll 1$. Noise-floor estimates show that Search Coil Magnetometer and Electric Double Probe sensitivity convergence produces a spurious apparent equipartition throughout this regime. The observed equality likely reflects nonlinear dynamics, incoherent superposition of electromagnetic and electrostatic fluctuations, or instrumental noise contamination.
\end{abstract}

\maketitle

%-----------------------------------------------------------------------
%\section{Introduction}

In collisionless plasma turbulence, the electric field mediates the transfer of energy from large-scale magnetic and flow fluctuations to particle kinetic energy at small scales~\cite{chasapis2018,Howes2024}. Understanding the statistical and spectral properties of the electric field in the kinetic range is therefore central to the problem of turbulent dissipation~\cite{Marino2023}.

Using MMS data from turbulent reconnection events in the Earth's magnetotail, Vo \textit{et~al.}~\cite{vo2026} recently reported that the electric and magnetic field energy spectral densities, $\varepsilon_0 P[\delta E]/2$ and $P[\delta B]/(2\mu_0)$, become approximately equal at frequencies above the electron gyroscale $f_{\rho_e}$. This electromagnetic energy equipartition was interpreted as evidence that the turbulent energy transfer mediated by the electric field reaches completion at sub-electron scales, resulting in relaxation to a state resembling thermodynamic equilibrium~\cite{Pathria2011}. Vo \textit{et~al.}~\cite{vo2026} acknowledge potential noise contamination in their sub-electron regime and flag affected spectral intervals with dotted lines; however, as we demonstrate quantitatively below, the transition from a physically-dominated to a noise-dominated electric field signal begins well below $f_{\rho_e}$, earlier than this flagging suggests.

In this paper, we examine whether the observed energy equipartition can arise from the linear polarization properties of wave modes known to exist in the kinetic range, specifically kinetic Alfv\'{e}n waves (KAWs) and whistler-mode waves. Gyrokinetic simulations spanning ion to electron scales have shown that linear KAW polarization relations quantitatively describe the turbulent fluctuation statistics even in a strongly nonlinear cascade~\cite{Howes2011}, which motivates the use of linear mode polarization as a baseline. We show analytically that for any non-relativistic plasma ($V_A \ll c$), the electromagnetic energy ratio predicted by linear wave theory remains well below unity at all wavenumbers, including the sub-electron range, placing a firm constraint on the interpretation of the observed equipartition.

%-----------------------------------------------------------------------
%\section{Electric-to-magnetic energy ratio from linear wave polarization}

We consider a uniform plasma with background magnetic field $\mathbf{B}_{0}=B_{0}\hat{z}$, ion (electron) mass $m_{i}$ ($m_{e}$), number density $n_{0}$, and temperatures $T_{i}$, $T_{e}$. The Alfv\'en speed is $V_{A}=B_{0}/\sqrt{\mu_{0}n_{0}m_{i}}$. Using the convention $v_{\mathrm{th},\alpha}^{2}=k_{\mathrm{B}}T_{\alpha}/m_{\alpha}$ for species $\alpha$, we define the ion (electron) gyroradius $\rho_{i}=v_{\mathrm{th},i}/\Omega_{i}$ ($\rho_{e}=v_{\mathrm{th},e}/\Omega_{e}$), the ion-sound gyroradius $\rho_{s}=c_{s}/\Omega_{i}$ with $c_{s}=\sqrt{k_{\mathrm{B}}T_{e}/m_{i}}$, the electron inertial length $d_{e}=c/\omega_{pe}$, and the electron plasma beta $\beta_{e}=2\mu_{0}n_{0}k_{\mathrm{B}}T_{e}/B_{0}^{2}$. The temperature ratio is $\tau=T_{i}/T_{e}$.

We define the electromagnetic energy spectral density ratio as
\begin{equation}
\mathcal{R}(k_\perp) \equiv \frac{\varepsilon_0 |\delta E|^2}{|\delta B|^2/\mu_0} = \frac{|\delta E|^2}{c^2 |\delta B|^2},
\label{eq:R_def}
\end{equation}
where $|\delta E|^2$ and $|\delta B|^2$ denote the total (all three components) mean-square field amplitudes of a linear eigenmode at perpendicular wavenumber $k_\perp$. The condition $\mathcal{R} = 1$ corresponds to equal electric and magnetic energy spectral densities.

%-----------------------------------------------------------------------
%\subsection{Kinetic Alfv\'{e}n waves}

The two-fluid KAW dispersion relation for a plane wave with wavevector $\mathbf{k} = k_\perp \hat{x} + k_\parallel \hat{z}$ is~\cite{Lysak1996,Stasiewicz2000,Hollweg1999}
\begin{equation}
\omega^2 = k_\parallel^2 V_A^2 \frac{1 + k_\perp^2 \rho_s^2}{1 + k_\perp^2 d_e^2}.
\label{eq:KAW_disp}
\end{equation}
This relation describes the shear Alfv\'{e}n mode from MHD scales ($k_\perp \rho_i \ll 1$) through the KAW regime ($k_\perp \rho_i \gg 1$, $k_\perp d_e \ll 1$) and into the inertial KAW regime ($k_\perp d_e \gg 1$). Ion finite-Larmor-radius corrections modify this dispersion at $\mathcal{O}(1)$ level near $k_{\perp}\rho_{i}\sim 1$ but do not alter the saturation $\mathcal{R}_{\infty}$, which depends only on $\rho_{s}^{2}/d_{e}^{2}$.

For the shear KAW eigenmode, the self-consistent solution of the two-fluid equations (Faraday's law, the electron and ion momentum equations, and quasineutrality) yields the standard polarization relation~\cite{Lysak1996,Hollweg1999}
\begin{equation}
\frac{\delta E_{\perp}}{\delta B_{\perp}} = \frac{\omega}{k_{\parallel}},
\label{eq:polarization}
\end{equation}
where $\delta E_{\perp}\equiv\delta E_{x}$ and $\delta B_{\perp}\equiv\delta B_{y}$ to leading order in $k_{\parallel}/k_{\perp}$. The parallel electric field satisfies $|\delta E_{\parallel}|/|\delta E_{\perp}| \sim (k_{\parallel}/k_{\perp})\,k_{\perp}^{2}\rho_{s}^{2}/(1+k_{\perp}^{2}\rho_{s}^{2})$, which follows from the isothermal-electron parallel momentum equation and quasineutrality~\cite{Boldyrev2013,Chen2013}. Its contribution to the total electric-field energy is therefore suppressed by $(k_{\parallel}/k_{\perp})^{2}$ relative to $|\delta E_{\perp}|^{2}$. The compressive magnetic perturbation $|\delta B_{\parallel}|^{2}$ is similarly subdominant at the kinetic scales of interest~\cite{Chen2013}. To leading order we therefore have $|\delta E|^{2}\simeq|\delta E_{\perp}|^{2}$ and $|\delta B|^{2}\simeq|\delta B_{\perp}|^{2}$, so that
\begin{equation*}
\mathcal{R}_\mathrm{KAW}(k_\perp) = \frac{1}{c^{2}}\!\left(\frac{\omega}{k_{\parallel}}\right)^{\!2} \!\left[1+\mathcal{O}\!\left(\tfrac{k_{\parallel}^{2}}{k_{\perp}^{2}}\right)\right].
\end{equation*}
Substituting the dispersion relation~\eqref{eq:KAW_disp} gives the KAW energy ratio
\begin{equation}
\mathcal{R}_\mathrm{KAW}(k_\perp) = \left(\frac{V_{A}}{c}\right)^{\!2} \frac{1+k_\perp^2\rho_s^2}{1+k_\perp^2 d_e^2}.
\label{eq:R_KAW}
\end{equation}
This expression has three asymptotic regimes.

\noindent\textit{(i) MHD scales} ($k_\perp \rho_s \ll 1$, $k_\perp d_e \ll 1$):
\begin{equation}
\mathcal{R}_\mathrm{KAW} \approx \frac{V_A^2}{c^2} \ll 1.
\label{eq:R_MHD}
\end{equation}

\noindent\textit{(ii) Sub-ion / KAW regime} ($k_\perp \rho_s \gg 1$, $k_\perp d_e \ll 1$):
\begin{equation}
\mathcal{R}_\mathrm{KAW} \approx \frac{V_A^2}{c^2}\,k_\perp^2 \rho_s^2.
\label{eq:R_subion}
\end{equation}
The ratio grows as $k_\perp^2$, reflecting the flattening of the electric field spectrum relative to the magnetic spectrum due to the Hall electric field contribution.

\noindent\textit{(iii) Sub-electron / inertial KAW regime} ($k_\perp \rho_s \gg 1$, $k_\perp d_e \gg 1$):
\begin{equation}
\mathcal{R}_\mathrm{KAW} \to \mathcal{R}_\infty \equiv \frac{V_A^2}{c^2}\frac{\rho_s^2}{d_e^2} = \frac{V_A^2}{c^2}\frac{m_i}{m_e}\frac{\beta_e}{2},
\label{eq:R_inf}
\end{equation}
where we used $\rho_s^2/d_e^2 = (m_i/m_e)(\beta_e/2)$, which follows from $\rho_s^2 = k_B T_e/(m_i \Omega_i^2)$ and $d_e^2 = c^2/\omega_{pe}^2$. The ratio saturates at a constant value determined solely by $V_A/c$, the mass ratio, and $\beta_e$.

Equation~\eqref{eq:R_inf} is a structural bound: the saturation value $\mathcal{R}_{\infty}$ depends only on the geometric ratio $\rho_{s}^{2}/d_{e}^{2}=(m_{i}/m_{e})(\beta_{e}/2)$ together with the overall $(V_{A}/c)^{2}$ prefactor. Since $V_{A}^{2}/c^{2}=\Omega_{i}^{2}/\omega_{pi}^{2}\ll 1$ in all non-relativistic heliospheric and magnetospheric plasmas, $\mathcal{R}_{\infty}$ remains well below unity despite the large mass ratio $m_{i}/m_{e}=1836$.

Fully kinetic (Vlasov--Maxwell) calculations show that electron thermal properties have non-trivial effects on the polarization of Alfv\'{e}nic modes at kinetic scales~\cite{VillarroelSepulveda2024,verscharen2022}, including a role for electron inertia in the warm-plasma regime. These corrections modify the detailed shape of $\mathcal{R}(k_\perp)$ near the electron gyroradius but \emph{cannot alter the fundamental non-relativistic ordering} that controls the overall magnitude of $\mathcal{R}$. The non-relativistic ordering $(V_A/c)^2 \ll 1$ is a constraint of Maxwell's equations themselves: $\varepsilon_0 |\delta E|^2$ is suppressed relative to $|\delta B|^2/\mu_0$ by $(v_\varphi/c)^2$, and no kinetic correction can make the wave phase speed $v_\varphi \sim V_A\sqrt{m_i \beta_e / 2m_e}$ approach $c$ in a non-relativistic plasma. Bridging the $\sim$500-fold gap between $\mathcal{R}_\infty$ and the observed $\mathcal{R} \sim 1$ would require a kinetic enhancement factor of order $500$ in $\mathcal{R}$, which would in turn require $v_\varphi \sim 22\,V_A\sqrt{m_i\beta_e/2m_e}$, a phase speed that would violate the non-relativistic assumption for any known heliospheric plasma. The saturation value $\mathcal{R}_\infty$ therefore remains a reliable order-of-magnitude upper bound on the electromagnetic energy ratio for any non-relativistic plasma, regardless of the fluid or kinetic framework used.

%-----------------------------------------------------------------------
%\subsection{Whistler-mode waves}

Whistler-mode waves are often invoked at sub-ion and sub-electron scales in reconnection-driven turbulence~\cite{Narita2016,Stawarz2015}. In the low-frequency limit ($\omega \ll \Omega_e$, $kd_e \ll 1$), the dispersion relation for whistler waves propagating at angle $\theta$ to $\mathbf{B}_0$ is~\cite{Stix1992}
\begin{equation}
\omega = k^2 d_e^2 \Omega_e \cos\theta.
\label{eq:whistler_disp}
\end{equation}
The phase speed $v_\varphi = \omega/k = kd_e^2 \Omega_e \cos\theta$ increases with $k$, and $|\delta E|/|\delta B| = v_\varphi$, so the whistler energy ratio is
\begin{equation}
\mathcal{R}_W(k) = \frac{k^2 d_e^4 \Omega_e^2 \cos^2\theta}{c^2}.
\label{eq:R_W}
\end{equation}
At $k\rho_e = 1$ (i.e., $k = \Omega_e/v_{\mathrm{th},e}$), this becomes
\begin{equation*}
\mathcal{R}_W\big|_{k\rho_e=1} = \frac{d_e^4 \Omega_e^4 \cos^2\theta}{v_{\mathrm{th},e}^2\,c^2} = \frac{V_A^4 m_i^2 \cos^2\theta}{m_e\,k_B T_e\,c^2},
\end{equation*}
using $d_e^2 \Omega_e^2 = V_A^2 m_i/m_e$. Although the fourth-power dependence on $V_A$ makes this smaller than $\mathcal{R}_\infty$ for most parameters, it remains of the same order of magnitude as the KAW result for typical magnetotail conditions.

\begin{figure}[ht]
\centering
\includegraphics[width=0.85\linewidth]{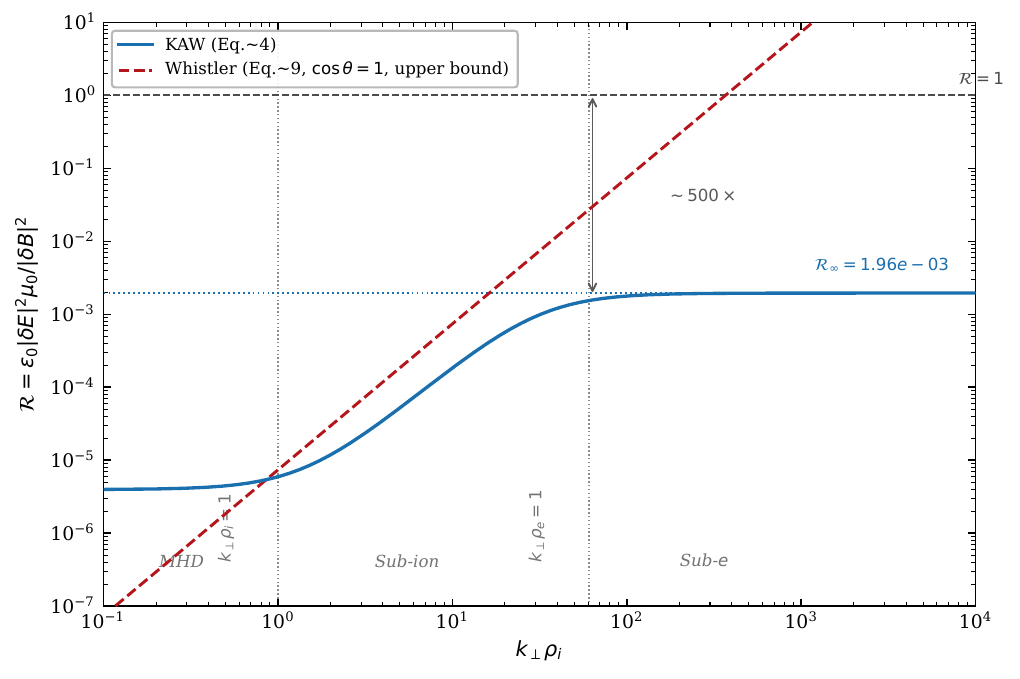}
\caption{Electromagnetic energy ratio $\mathcal{R}$ versus normalized perpendicular wavenumber $k_{\perp}\rho_{i}$ for kinetic Alfv\'en waves [Eq.~\eqref{eq:R_KAW}, solid blue] and whistler-mode waves [Eq.~\eqref{eq:R_W} with $\cos\theta=1$, red; upper bound on $\mathcal{R}_{W}$]. The whistler curve is solid within its formal validity domain ($kd_{e}\le 1$) and dashed beyond it, where the low-frequency cold-plasma dispersion relation no longer applies. Regimes are labeled MHD ($k_{\perp}\rho_{i}\lesssim 1$), Sub-ion ($1\lesssim k_{\perp}\rho_{i}\lesssim \rho_{i}/\rho_{e}$), and Sub-electron ($k_{\perp}\rho_{e}\gtrsim 1$). Plasma parameters are representative of the MMS magnetotail reconnection events of Vo \textit{et~al.}~\cite{vo2026}. The KAW ratio saturates at $\mathcal{R}_{\infty}\approx 1.96\times 10^{-3}$, approximately 500 times below equipartition ($\mathcal{R}=1$, horizontal dashed line); within the whistler validity range, $\mathcal{R}_{W}\le 7.3\times 10^{-3}$.}
\label{fig:R_vs_k}
\end{figure}

The low-frequency dispersion relation~\eqref{eq:whistler_disp} requires $kd_e \ll 1$. For the magnetotail plasma parameters adopted here, $d_e/\rho_e \approx 1.93$, so this condition is already marginal at $k\rho_e \sim 1$ ($kd_e \approx 1.93$) and is violated for $k\rho_e \gtrsim 1$. This limitation, however, \emph{strengthens} rather than weakens our conclusion: within the strictly valid domain ($kd_e \leq 1$, $k_\perp \rho_i \lesssim 31$), the whistler ratio already reaches at most $\mathcal{R}_W \leq 7.3\times10^{-3}$, nearly two orders of magnitude below equipartition. For the whistler mode to produce $\mathcal{R} \sim 1$ at sub-electron scales, the ratio would need to increase by a factor of $\sim\!140$ within a single decade of wavenumber beyond the formal validity range: a requirement for which there is no supporting theoretical mechanism. In Fig.~\ref{fig:R_vs_k}, the whistler curve is plotted with $\cos\theta = 1$ (an upper bound on $\mathcal{R}_W$) as a solid line within the formal validity domain and as a dashed extension beyond it. We note that the cold-fluid dispersion relation used here tends to underestimate $|\delta E|/|\delta B|$ for oblique whistler waves at $\beta \sim 1$ relative to hot-plasma kinetic calculations~\cite{Huang2019}; our $\mathcal{R}_W$ is therefore likely a conservative lower bound on the true kinetic whistler ratio at $k\rho_e \sim 1$. Even accounting for a kinetic correction factor of a few, the whistler ratio remains well below unity for the parameters considered here, so the main conclusion is unaffected.

%-----------------------------------------------------------------------
%\section{Application to magnetotail reconnection turbulence}

We now evaluate $\mathcal{R}(k_\perp)$ using plasma parameters representative of the reconnection-driven turbulence events studied by Vo \textit{et~al.}~\cite{vo2026}. Based on typical MMS magnetotail observations near the ion diffusion region~\cite{Ergun2018,Vo2023}, we adopt $B_0 = 15\,\mathrm{nT}$, $n_0 = 0.3\,\mathrm{cm}^{-3}$, $T_i = 2\,\mathrm{keV}$, $T_e = 1\,\mathrm{keV}$, yielding:
\begin{align*}
V_A &\approx 598\;\mathrm{km/s}, \quad V_A/c \approx 2.0\times10^{-3}, \\
\beta_i &\approx 1.07, \quad \beta_e \approx 0.54, \quad \tau = T_i/T_e = 2, \\
\rho_i &\approx 305\;\mathrm{km}, \quad \rho_e \approx 5.0\;\mathrm{km}, \\
\rho_s &\approx 216\;\mathrm{km}, \quad d_e \approx 9.7\;\mathrm{km}, \quad d_e/\rho_e \approx 1.93.
\end{align*}
The asymptotic KAW energy ratio from Eq.~\eqref{eq:R_inf} is
\begin{equation*}
\mathcal{R}_\infty = (2.0\times10^{-3})^2 \times 1836 \times \frac{0.54}{2} \approx 2.0\times10^{-3},
\end{equation*}
approximately 500 times smaller than the equipartition value $\mathcal{R} = 1$ reported by Vo \textit{et~al.}~\cite{vo2026}. Direct evaluation at $k_\perp \rho_e = 1$ gives $\mathcal{R}_\mathrm{KAW} \approx 1.5\times10^{-3}$ (approximately 667 times below equipartition) and $\mathcal{R}_W \approx 2.7\times10^{-2}$ with $\cos\theta = 1$ (approximately 37 times below equipartition). Within the strictly valid whistler range ($kd_e \leq 1$, $k_\perp \rho_e \lesssim 0.5$), the whistler ratio reaches at most $\mathcal{R}_W \leq 7.3\times10^{-3}$. Both wave modes predict $\mathcal{R} \ll 1$ throughout the sub-electron range.

\begin{figure}[ht]
\centering
\includegraphics[width=0.85\linewidth]{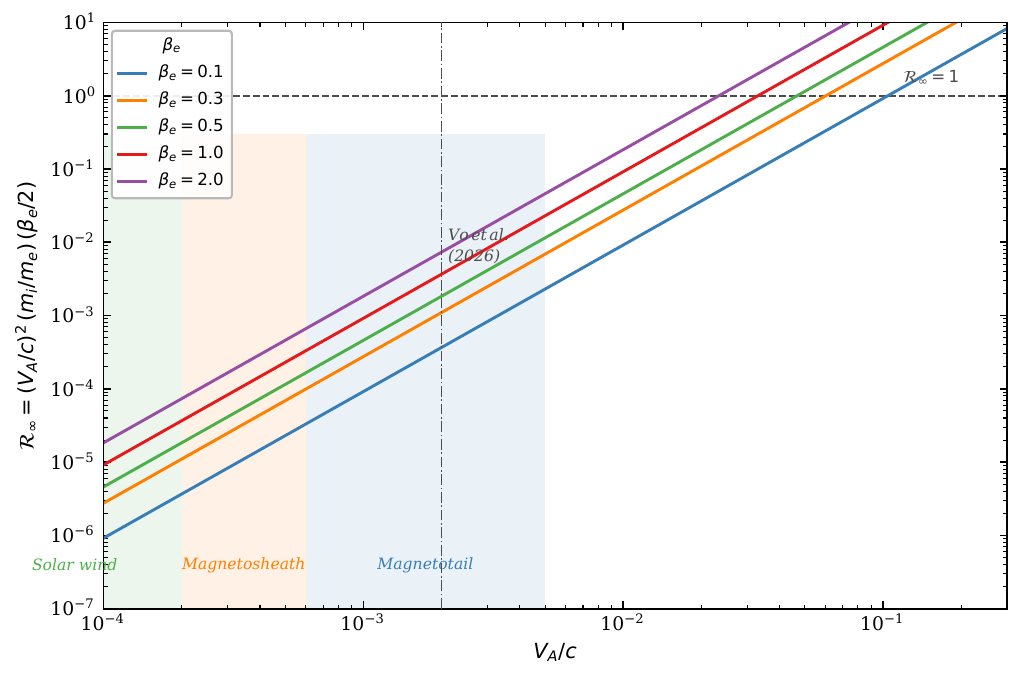}
\caption{Asymptotic KAW energy ratio $\mathcal{R}_\infty=(V_A/c)^2(m_i/m_e)(\beta_e/2)$ as a function of $V_A/c$ for several values of $\beta_e$. Shaded bands indicate typical $V_A/c$ ranges for the solar wind, magnetosheath, and magnetotail. Equipartition ($\mathcal{R}_\infty=1$) requires $V_A/c\gtrsim0.033$ at $\beta_e\sim1$, far beyond any non-relativistic heliospheric condition.}
\label{fig:R_vs_VA}
\end{figure}

The full wavenumber dependence $\mathcal{R}_\mathrm{KAW}(k_\perp)$ from Eq.~\eqref{eq:R_KAW} is plotted in Fig.~\ref{fig:R_vs_k}, along with $\mathcal{R}_W$ from Eq.~\eqref{eq:R_W}. Figure~\ref{fig:R_vs_VA} shows $\mathcal{R}_\infty$ as a function of $V_A/c$ for several values of $\beta_e$. Setting $\mathcal{R}_\infty = 1$ gives the general equipartition threshold
\begin{equation}
\frac{V_A}{c} \gtrsim \sqrt{\frac{2}{(m_i/m_e)\,\beta_e}},
\label{eq:threshold}
\end{equation}
which evaluates to $V_A/c \gtrsim 0.033$ ($V_A \gtrsim 9{,}900\,\mathrm{km/s}$) for $\beta_e \sim 1$, and $V_A/c \gtrsim 0.023$ ($V_A \gtrsim 7{,}000\,\mathrm{km/s}$) for $\beta_e \sim 2$. Neither condition is met in any non-relativistic heliospheric or magnetospheric plasma, where $V_A/c \sim 10^{-4}$--$10^{-2}$.

The result holds generally: for any non-relativistic plasma with $V_A/c \ll 1$,
\begin{equation*}
\mathcal{R}_\infty = \frac{V_A^2}{c^2}\frac{m_i}{m_e}\frac{\beta_e}{2} \ll 1,
\end{equation*}
irrespective of the temperature ratio, the spectral anisotropy, or the specific details of the turbulent cascade.

%-----------------------------------------------------------------------
%\section{Discussion}

The linear polarization of KAWs and whistler waves in a non-relativistic plasma predicts $\mathcal{R} \lesssim 10^{-3}$ at all heliospheric parameters and all perpendicular wavenumbers through the sub-electron range. This follows from the non-relativistic ordering: $\varepsilon_0|\delta E|^2/2$ is suppressed relative to $|\delta B|^2/(2\mu_0)$ by $(v_\varphi/c)^2$, with $v_\varphi = V_A\sqrt{m_i\beta_e/2m_e} \ll c$ for KAWs.

The observation of $\mathcal{R} \sim 1$ in the sub-electron range~\cite{vo2026} therefore cannot be a signature of linear wave dynamics. We identify three possible origins, in decreasing order of parsimony given the present evidence.

\textit{(i) Instrument noise convergence.} The MMS Search Coil Magnetometer (SCM) operates in burst mode with a noise amplitude spectral density of $\approx 10^{-5}\,\mathrm{nT}/\sqrt{\mathrm{Hz}}$ for $f \gtrsim 8\,\mathrm{Hz}$~\cite{LeContel2016}. At the relevant (field-aligned) frequencies, the dominant electric-field noise source is the Axial Double Probe (ADP)~\cite{Ergun2016}, with a noise floor of $\approx 0.3\,\mathrm{mV/m}/\sqrt{\mathrm{Hz}}$ for $f \gtrsim 1\,\mathrm{Hz}$. Converting to energy spectral densities:
\begin{align}
\frac{P_\mathrm{noise}[\delta B]}{2\mu_0} &\sim \frac{(10^{-14}\,\mathrm{T})^2/\mathrm{Hz}}{2\mu_0} \sim 4\times10^{-23}\;\frac{\mathrm{J}}{\mathrm{m}^3\,\mathrm{Hz}}, \label{eq:SCM_noise}\\
\frac{\varepsilon_0 P_\mathrm{noise}[\delta E]}{2} &\sim \frac{\varepsilon_0(3\times10^{-4}\,\mathrm{V/m})^2/\mathrm{Hz}}{2} \sim 4\times10^{-19}\;\frac{\mathrm{J}}{\mathrm{m}^3\,\mathrm{Hz}}. \label{eq:ADP_noise}
\end{align}
The ADP noise energy density therefore exceeds the SCM noise energy density by approximately four orders of magnitude. This four-order-of-magnitude asymmetry is an instrumental property independent of plasma parameters, making the qualitative conclusion robust across a wide range of magnetotail conditions.

\begin{figure}[ht]
\centering
\includegraphics[width=0.85\linewidth]{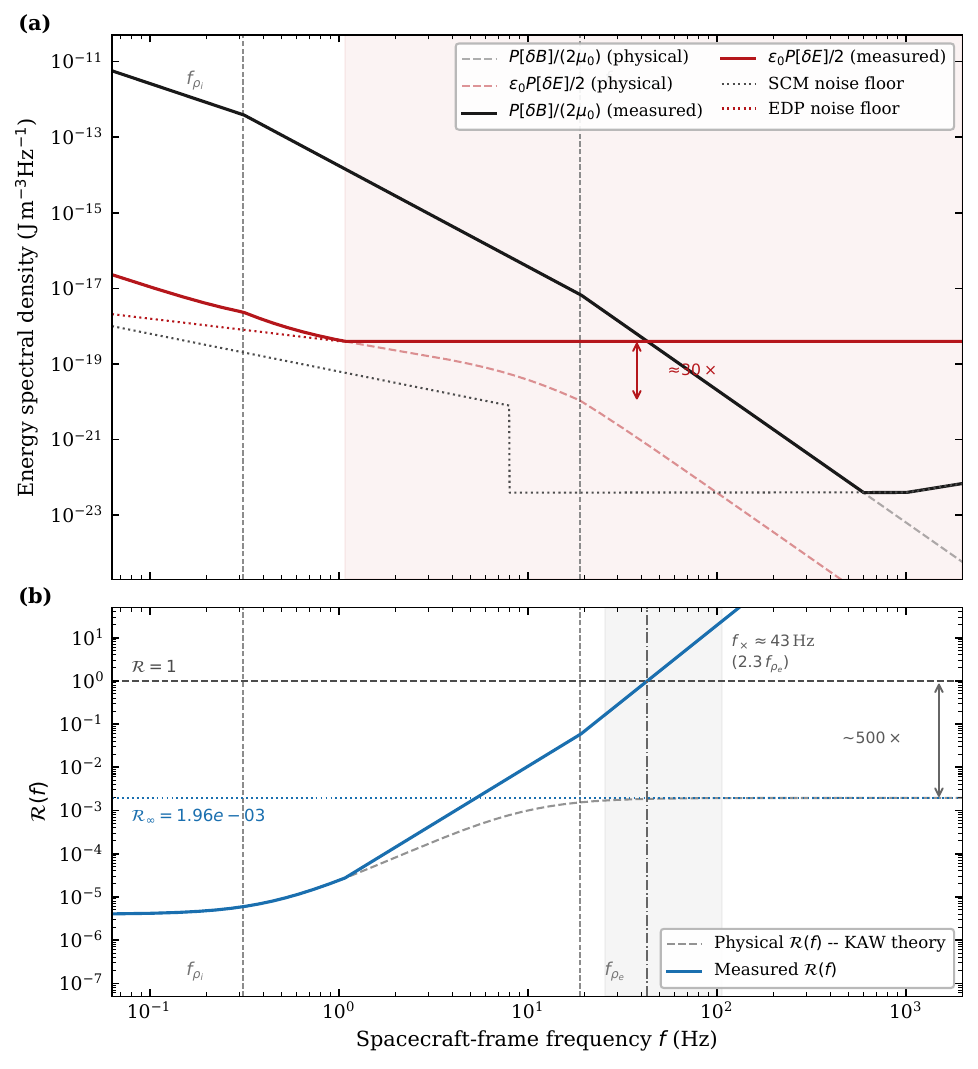}
\caption{MMS noise-floor analysis ($V_{\mathrm{str}}=V_{A}$). (a)~Energy spectral densities $P[\delta B]/(2\mu_{0})$ (black) and $\varepsilon_{0}P[\delta E]/2$ (red), in $\mathrm{J\,m^{-3}\,Hz^{-1}}$. Dashed curves show the physical KAW signal; solid curves show the measured signal including instrument noise. The physical $\delta E$ drops below the ADP noise floor at $f\approx 3.5\,f_{\rho_{i}}$, well before the electron gyroscale $f_{\rho_{e}}$ (shaded band marks the noise-dominated regime). (b)~Measured energy ratio $\mathcal{R}(f)$ rises spuriously and crosses $\mathcal{R}=1$ at $f\approx 43\,\mathrm{Hz}\approx 2.3\,f_{\rho_{e}}$, within regime IV of Vo \textit{et~al.}~\cite{vo2026}.}
\label{fig:noise}
\end{figure}

For the plasma parameters described above, the physical $\delta E$ spectral density is modeled using the KAW polarization relation (Eq.~\ref{eq:R_KAW}), anchored to the observed sub-ion power level from Vo \textit{et~al.}~\cite{vo2026} and propagated using the KAW spectral slopes of Eqs.~\eqref{eq:R_subion}--\eqref{eq:R_inf}. The physical signal exceeds the ADP noise floor by a factor of $\approx 30$ at $f_{\rho_i}$, but this margin is entirely consumed by $f \approx 1\,\mathrm{Hz} \approx 3.5\,f_{\rho_i}$, well below the electron gyroscale $f_{\rho_e} \approx 19\,\mathrm{Hz}$, while the physical magnetic field signal remains above the SCM noise floor at all relevant frequencies. As a consequence, the measured energy ratio $\mathcal{R}_\mathrm{meas}(f)$ rises spuriously and crosses $\mathcal{R} = 1$ at $f \approx 43\,\mathrm{Hz} \approx 2.3\,f_{\rho_e}$, precisely within regime IV of Vo \textit{et~al.}~\cite{vo2026}. This is illustrated in Fig.~\ref{fig:noise}. We note that Figure~\ref{fig:noise} presents a model-based noise floor estimate using representative parameters; while event-specific noise floor characterization would provide a more definitive test, the four-orders-of-magnitude gap between ADP and SCM noise energy densities (Eqs.~\ref{eq:SCM_noise}--\ref{eq:ADP_noise}) is an instrumental property independent of plasma parameters, making the qualitative conclusion reliable.

Increasing the Taylor-hypothesis structure speed to $V_\mathrm{str} = 3V_A$ (characteristic of fast reconnection outflow) shifts the electron gyroscale to $f_{\rho_e} \approx 57\,\mathrm{Hz}$ and moves the ADP noise crossover to $f \approx 3\,\mathrm{Hz}$; the qualitative conclusion is unchanged and the noise argument becomes \emph{more} pronounced, since $f_{\rho_e}$ moves further above the noise-dominated onset frequency. The Taylor hypothesis uncertainty therefore strengthens rather than weakens the noise floor interpretation. We caution, however, that the Taylor hypothesis may be strained in reconnection outflow where wave phase speeds are comparable to the bulk flow~\cite{Howes2014}, and event-specific assessment remains desirable.

Vo \textit{et~al.}~\cite{vo2026} acknowledge noise contamination in regime IV and flag affected frequencies with dotted lines. However, the analysis above shows that the ADP noise floor overtakes the physical $\delta E$ signal already at $f \approx 3.5\,f_{\rho_i}$, more than five times below $f_{\rho_e}$, so the contamination begins well before the spectral interval their flagging is intended to bracket. The measured $\mathcal{R} \sim 1$ in regime IV is therefore consistent with noise contamination of the electric field signal across the entire sub-electron frequency range. This mechanism alone is quantitatively sufficient to produce the observed $\mathcal{R} \sim 1$, requires no additional physical assumptions, and is directly testable.

\textit{(ii) Incoherent superposition of wave modes.} At sub-electron scales, the turbulence may contain a mixture of electromagnetic (KAW, whistler) and electrostatic modes (electron-acoustic waves, electron holes, Langmuir waves)~\cite{Ergun2009,Stawarz2015,Lotekar2020}. Electrostatic fluctuations contribute to $P[\delta E]$ without a corresponding $\delta B$ signature. For the total measured $\mathcal{R}$ to reach unity, electrostatic modes would need to carry roughly $500\times$ more electric field energy at sub-electron scales than the electromagnetic modes do, an extraordinary requirement whose observational status remains open. In this scenario, the apparent ``equipartition'' reflects coincidental matching of electric and magnetic energy budgets from distinct wave populations, not a property of any single mode.

\textit{(iii) Genuinely nonlinear dynamics.} Strong turbulence at sub-electron scales may generate coherent structures (current sheets, electron holes, double layers) whose $\delta E/\delta B$ ratios are not constrained by linear wave polarization. Localized regions of intense parallel electric field associated with electron-scale structures~\cite{Pathak2025} could contribute disproportionately to $P[\delta E]$ at high frequencies, as suggested by the non-Gaussian statistics reported by Vo \textit{et~al.}~\cite{vo2026}. A related caution is that fully kinetic PIC simulations of decaying turbulence at physical mass ratio suggest that identifying electron-scale turbulent fluctuations as a superposition of linear modes is not straightforward~\cite{Camporeale2011}. Whether this manifests as an enhancement of $\mathcal{R}$ at sub-electron scales remains to be examined quantitatively, but it does suggest that the linear baseline we derive should be understood as a reference rather than a strict bound at $k_\perp \rho_e \gtrsim 1$. For nonlinear structures alone to account for $\mathcal{R} \sim 1$ in the absence of noise, they would still need to collectively bridge the large $\sim\!500\times$ gap relative to the two-fluid prediction, requiring an exceptionally high filling fraction of strongly driven, high-$\delta E/\delta B$ structures throughout the sub-electron range.

We note that the scaling $\alpha_E = \alpha_B - 2$ in the sub-ion kinetic range, observed by Vo \textit{et~al.}~\cite{vo2026} and predicted by Hall-MHD turbulence~\cite{Alexandrova2008}, is fully consistent with the $k_\perp^2$ growth of $\mathcal{R}_\mathrm{KAW}$ in Eq.~\eqref{eq:R_subion}, confirming that linear KAW polarization adequately describes the relative $E$ and $B$ spectra in the sub-ion range. The departure from this scaling at $k_\perp \rho_e \sim 1$, where the spectra converge, coincides precisely with the onset of ADP noise dominance in the electric field measurement.

%-----------------------------------------------------------------------
%\section{Conclusions}

We have shown analytically that the electromagnetic energy ratio $\mathcal{R}(k_\perp) = \varepsilon_0|\delta E|^2/(|\delta B|^2/\mu_0)$ predicted by the linear two-fluid dispersion relations of kinetic Alfv\'{e}n waves and whistler waves satisfies
\begin{equation}
\mathcal{R}(k_\perp) \leq \frac{V_A^2}{c^2}\frac{m_i}{m_e}\frac{\beta_e}{2} \ll 1
\label{eq:R_bound}
\end{equation}
at all perpendicular wavenumbers in any non-relativistic plasma. For typical magnetotail parameters, the two-fluid upper bound is $\mathcal{R}_\infty \approx 2\times10^{-3}$, approximately 500 times below the apparent energy equipartition $\mathcal{R} \sim 1$ reported by Vo \textit{et~al.}~\cite{vo2026}. Achieving equipartition from linear wave polarization alone requires $V_A/c \gtrsim 0.033$ for $\beta_e \sim 1$, corresponding to $V_A \gtrsim 10{,}000\,\mathrm{km/s}$, a condition satisfied in no known non-relativistic heliospheric or magnetospheric environment.

This $\sim$500-fold gap is governed by the non-relativistic ordering $(V_A/c)^2 \ll 1$, which is a constraint of Maxwell's equations independent of the fluid or kinetic framework. Electron FLR and other kinetic corrections at $k_\perp \rho_e \sim 1$ modify the detailed shape of $\mathcal{R}(k_\perp)$ but cannot change its fundamental scale: making $\varepsilon_0|\delta E|^2 \sim |\delta B|^2/\mu_0$ requires a wave phase speed $v_\varphi \sim c$, which is not achievable in a non-relativistic plasma by any known kinetic correction mechanism.

This result suggests that the observed electromagnetic energy equipartition at sub-electron scales is not a consequence of linear wave polarization alone, and most plausibly originates from nonlinear dynamics, incoherent superposition of electromagnetic and electrostatic fluctuations, or convergence of signals toward instrument noise floors. The MMS noise-floor analysis presented here shows that the third mechanism alone is quantitatively sufficient to produce $\mathcal{R}_\mathrm{meas} \approx 1$ in regime IV, with the physical electric field signal falling below the ADP noise floor at $f \approx 3.5\,f_{\rho_i}$, well before the onset of the sub-electron regime. The four-orders-of-magnitude asymmetry between ADP and SCM noise energy densities is an instrumental property independent of plasma parameters, and the Taylor-hypothesis uncertainty only strengthens the noise interpretation at higher structure speeds.

A decisive observational test would be to identify events in the MMS magnetotail dataset where the physical $\delta E$ signal is demonstrably above the ADP noise floor at $f > f_{\rho_e}$ (for example, in high-$\beta_e$ or high-density events where $\mathcal{R}_\infty$ is larger and the physical signal is stronger relative to the noise floor) and to check whether $\mathcal{R} \sim 1$ persists; if it does, instrumental noise can be ruled out as the primary cause. A complementary multi-component test uses the ratio $P[\delta E_{\parallel}]/P[\delta E_{\perp}]$ above $f_{\rho_{e}}$: noise-dominated intervals drive this ratio toward unity as both components sit on the ADP floor, whereas linear KAW dynamics predict $P[\delta E_{\parallel}]/P[\delta E_{\perp}]\lesssim (k_{\parallel}/k_{\perp})^{2}\ll 1$. The two scenarios are therefore observationally separable in MMS burst-mode data. Distinguishing among the three possible origins ultimately requires per-event noise-floor characterization combined with multi-point wave-mode identification at sub-electron scales.

%-----------------------------------------------------------------------
%-----------------------------------------------------------------------

%%%%%%%%%%%%%%%%%%fig 2

%fig 3

\paragraph*{Data availability.}%
The analytical results are fully reproducible from Eqs.~\eqref{eq:KAW_disp}--\eqref{eq:threshold}. The noise-floor model uses published Search Coil Magnetometer and Axial Double Probe instrument specifications from Refs.~\cite{LeContel2016,Ergun2016}. The plasma parameters are drawn from the published MMS magnetotail observations cited in the text. No new observational data were generated in this study.

\begin{acknowledgments}

MKC and BS acknowledge the University Grants Commission (UGC), India, for support through the Non-NET Fellowship programme. One of the authors RM acknowledges support from IUCAA Pune through visiting associate program.

\end{acknowledgments}

\bibliography{reference}

\end{document}